\documentclass[%
preprint,
 amsmath,amssymb,
 aps,
]{revtex4-2}

\usepackage{graphicx}
\usepackage{dcolumn}
\usepackage{bm}
\usepackage{color}

\begin{document}

\preprint{preprint}

\title{
A Model without Higgs Potential for Quantum Simulation of Radiative Mass-Enhancement in SUSY Breaking}%

\author{Masao Hirokawa}
\email{hirokawa@inf.kyushu-u.ac.jp}
\homepage{https://nvcspm.net/qstl/}
\altaffiliation{Graduate School of Information Science and Electrical Engineering, Kyushu University.}

\date{\today}

\begin{abstract}
  We study a quantum-simulation model of a mass enhancement
  in the fermionic states, as well as in the bosonic ones, 
  of the supersymmetric quantum mechanics.
  The bosonic and fermionic states are graded by a qubit. 
  This model is so simple that it may be implemented as
  a quantum simulation of the mass enhancement taking place 
  when supersymmetry (SUSY) is spontaneously broken.
  Here, our quantum simulation means
  the realization of the target quantum phenomenon
  with some quantum-information devices as a physical reality.
  The model describes how the quasi-particle consisting of
  the annihilation and creation of 1-mode scalar bosons 
  eats the spin effect given by the $X$-gate,
  and how it acquires the mass enhancement
  in the fermionic states in the spontaneous SUSY breaking.
  Our model's interaction does not have any Higgs potential.
  Instead, the qubit acts as a substitute for the Higgs potential
  by the 2-level-system approximation of the double-well potential,
  and then, the spontaneous SUSY breaking takes place and
  the mass is enhanced.
\end{abstract}

\maketitle


\section{Introduction}\label{sec1}

In 2012 the long-sought Higgs boson is found \cite{atl12, cms12}, which establishes
the triumph of the Brout-Englert-Higgs mechanism \cite{eng64, hig64}.
This mechanism tells us how no-mass gauge particles gain mass in the standard model (SM),
while the gauge particle itself alone cannot have its mass due to gauge symmetry.
That finding shows the Higgs-particle mass of $125$ GeV ($\sim 10^{2}$ GeV).
When considering the interaction of the Higgs particle
in the theory including both the electroweak scale and the Planck scale,
particle physicists normally need a special tuning to obtain the Higgs mass \cite{sus79}.
Since the Planck-scale mass ($\sim 10^{18}$ GeV) is so much heavier than the Higgs mass,
particle physicists usually employ the so-called fine-tuning in SM
to cope with the mass gap with the ratio ($\sim 10^{16}$ GeV); thus,
they perform the unnatural, huge cancellation between the bare mass term
and the quantum corrections
to obtain the Higgs mass.
This is the so-called hierarchy problem. 
Moreover, the Higgs mass of $125$ GeV could result in the possibility of
the flat Higgs potential
when the electroweak scale is together with the Planck scale
\cite{hol12, eli12, deg12, but13, iso13, ibe14}.
It says that the Higgs quartic interaction may be invalidated.
Removing this apprehension, we probably should need to find a mass-enhancement mechanism
by the radiative generation without the Higgs potential. 
Against these difficulties, supersymmetry (SUSY)
is among the strong candidates
for natural theories to solve those problems. 
However, the Higgs mass of $125$ GeV puzzles particle physicists again because it is
rather heavier than the mass predicted in the minimal supersymmetric standard model (MSSM).
The mass of $125$ GeV is almost the upper bound ($110-135$ GeV) of
the possibly predicted mass,
and imposes pretty tight constraints on the conditions of MSSM \cite{arb12}. 
This gap between the two masses requires another fine-tuning. 
It is expected that this gap is plugged
by the SUSY breaking \cite{sal74, buc82, giu99, arb12,
  dra12, dud13, iba13, ant14, lu14, oku19}, a kind of spontaneous symmetry breaking.

In the light of relativistic quantum field theory,
although Coleman and Mandula's no-go theorem states
non-trivial theory's impossibility of
combining the Poincar\'{e} symmetry and internal one
\cite{cm67},
the Haag-{\L}opusza\'{n}ski-Sohnius theorem
gives us a loophole in the Coleman-Mandula theorem,
which says that a way nontrivially to mix the Poincar\'{e} and internal
symmetries is through SUSY \cite{hls75}.
Excluding conformal field theory, 
SUSY may be the last bastion for the combination of
the Poincar\'{e} and internal symmetries in relativistic quantum field theory.
In other words, if SUSY is not a physical reality,
the two theorems show the theoretical limitation
of relativistic quantum field theory. 

Unfortunately, any superpartner (i.e., supersymmetric particle paired
with an elementary particle) has not yet been found \cite{atl21};
in fact, any fingerprint of SUSY and its spontaneous breaking
had not been firmly, directly observed.
We note that a vestige of SUSY is found in atomic nuclei \cite{met99}.
We probably should study what parts of the theory of SUSY are realized as a physical reality,
and clarify them one by one.
In the first place, we should confirm the physically real existences of
SUSY and its spontaneous breaking. 
Witten squeezes the minimal essence of the supersymmetric quantum mechanics (SUSY QM),
and develops it in quantum mechanics \cite{wit81, wit82}.
Although the verification of the full theory of SUSY needs a huge facility,
that for SUSY QM \cite{bin06, gmr11, bau15a, bau15b} requires the reasonable facility in a laboratory. 
Actually, Cai \textit{et al.} report that
they succeed in observing  \cite{cai22} a signature of SUSY QM.

Some months before the Higgs-boson discovery, actually, the quantum simulation for
the Brout-Englert-Higgs mechanism is succeeded \cite{end12}.
Quantum simulation is for the study of quantum phenomena,
and is implemented on a programmable quantum system
consisting of quantum devices especially designed to realize those quantum phenomena.
In other words, it realizes the target quantum phenomenon
appearing in physics such as the elementary particles theory 
with using some physics in a laboratory as a physical reality.
Therefore, the quantum simulation is different from the virtual simulations by 
conventional computers. 
The original idea of quantum simulation is based on Feynman's proposal \cite{fey82}
and has experimentally been developed\cite{ger11,end12,yan16,est16,kok19,sch19,yan20,cai22,zha22}.
Some theoretical models for quantum simulation of SUSY and its spontaneous breaking
are proposed \cite{hir11,hir15,tom15,url15,gha21,min22}.
In particular, a simple prototype model is given,
and it has the transition from the $\mathcal{N}=2$\, SUSY to its spontaneous breaking \cite{hir11,hir15}.
It is based on the quantum Rabi model \cite{rab36, rab37, bra11}.
The quantum Rabi model is the $1$-mode scalar boson version of the spin-boson model \cite{leg87}. 
The success in an experimental observation is reported,
and that transition is observed in a trapped ion quantum simulator \cite{cai22}. 
In this transition we cannot observe any mass enhancement
in the fermionic states as well as in the bosonic ones
because the Lagrangian of the prototype model does not include
any mass-enhancement mechanism.
Thus, we are interested in quantum simulation showing
a mass enhancement in the fermionic states in SUSY breaking.
One of the candidates for the mass enhancement is adding the quadratic term,
often called `$A^{2}$-term' \cite{rza75, nat10},
for the bosonic states 
as an extra mass term to the quantum Rabi model.
Since the quantum Rabi model describes the electromagnetic interaction
basically,
its `$A$' corresponds to the photon gauge field. 
For the prototype model \cite{hir11,hir15}, the strong coupling limit is used to obtain the transition.
As shown in this paper, however, we can derive a no-go theorem for
the SUSY breaking in the strong coupling limit
if the prototype model has the $A^{2}$-term.
On the other hand, Cai \textit{et al.} propose another limit experimentally to obtain
the transition for the prototype model \cite{cai22}.
We show that their limit makes our model avoid the no-go theorem.
Employing their limit, therefore, we extend the prototype model 
such that we can make quantum simulation for the mass enhancement in
SUSY breaking.
Our model's interaction has no Higgs potential, and thus,
the mass enhancement of the bosonic states is radiatively made
by its 2-level-system approximation.
In other words, a qubit coupled with the 1-mode scalar boson
works as a substitute for the Higgs potential in our system.

In this paper we consider scalar boson only.
Thus, we call scalar boson merely ``boson'' for short.  
The structure of this paper is as follows:
In Section \ref{sec2} we prove that the quantum Rabi model with
the $A^{2}$-term meets the no-go theorem for the SUSY breaking
in the strong coupling limit.
On the other hand, we also prove that it can avoid the no-go theorem
under the scheme by Cai \textit{et al.} \cite{cai22}, 
and it has the transition from the $\mathcal{N}=2$\, SUSY to its spontaneous breaking.
In Section \ref{sec3} we show that the mass enhancement
in the fermionic states as well as in the bosonic ones takes place
in the SUSY breaking.
We explain what works for spontaneous symmetry breaking
in the mass-enhancement process instead of the Higgs potential. 
In Section \ref{sec4} we discuss the experimental realization
of our quantum simulation.
We introduce some problems on the Goldstino (i.e., Nambu-Goldstone fermion)
arising from the results in this paper.

\section{Quantum Rabi model in SUSY QM}\label{sec2}

In this section, we explain the role of the quantum Rabi model
for the transition from the $\mathcal{N}=2$\, SUSY to its spontaneous breaking.
The quantum Rabi model has been coming in handy for quantum simulation lately
\cite{bra17, yos17, lv18, cai21, mei22},
and it can be a powerful tool for our purpose.

The state space of the 1-mode boson is given by the boson Fock space $\mathcal{F}_{\mathrm{b}}$,
which is spanned by the boson Fock states.
The boson Fock state with $n$ bosons is denoted by $\mid\!\!n\rangle$;
thus, $\mid\!\!0\rangle$ is the Fock vacuum in particular.
The $2$-level atom in our model is represented by spin. 
We denote the up-spin state by
$\mid\uparrow\rangle=\bigl(
\begin{smallmatrix}
   1 \\
   0
\end{smallmatrix}
\bigl)$, 
and the down-spin state by
$\mid\downarrow\rangle=\bigl(
\begin{smallmatrix}
   0 \\
   1
\end{smallmatrix}
\bigl)$.
We denote by $\mathbb{C}$ the set of all the complex numbers. 
Then, $\mathbb{C}^{2}$ is the $2$-dimensional unitary space
with the natural inner product.
We use the Hilbert space $\mathbb{C}^{2}\otimes\mathcal{F}_{\mathrm{b}}$
for the total state space of our model.
The orthonormal basis of $\mathbb{C}^{2}\otimes\mathcal{F}_{\mathrm{b}}$
is given by the set
of all the vectors $\mid\downarrow\rangle\otimes\mid\!\!n\rangle$
and $\mid\uparrow\rangle\otimes\mid\!\!n'\rangle$ for
$n, n'=0, 1, 2, \cdots$.  
We often omit the symbol `$\otimes$'
in the vectors of $\mathbb{C}^{2}\otimes\mathcal{F}_{\mathrm{b}}$
throughout this paper.
The annihilation and creation operators
of a $1$-mode boson are respectively denoted by $a$ and $a^{\dagger}$.
The annihilation operator $\sigma_{-}$ and the creation operator
$\sigma_{+}$ of a $2$-level atom, that is, spin or qubit, are given by
$\sigma_{\pm}=(1/2)(\sigma_{x}\pm i\sigma_{y})$.
Thus, $\sigma_{-}$ and $\sigma_{+}$ are respectively the spin-annihilation operator 
and spin-creation operator.
Here, the standard notations, $\sigma_{x}$, $\sigma_{y}$, and
$\sigma_{z}$, are used for the Pauli matrices:
$\sigma_{x}=\bigl(
\begin{smallmatrix}
   0 & 1 \\
   1 & 0
\end{smallmatrix}
\bigl)$,
$\sigma_{y}=\bigl(
\begin{smallmatrix}
   0 & -i \\
   i & 0
\end{smallmatrix}
\bigl)$,
and 
$\sigma_{z}=\bigl(
\begin{smallmatrix}
   1 & 0 \\
   0 & -1
\end{smallmatrix}
\bigl)$.
We use the notation `$1$' for the 2-by-2 identity matrix, i.e.,
$1=\bigl(
\begin{smallmatrix}
   1 & 0 \\
   0 & 1
\end{smallmatrix}
\bigl)$,
and for the identity operator acting in $\mathcal{F}_{\mathrm{b}}$
as well as the numerical character $1$.
We often omit the symbols, `$1\otimes$' and `$\otimes 1$,'
in operators throughout this paper.

\subsection{Our problems}

We consider the physical system consisting of the $2$-level atom and $1$-mode boson. 
The two ideal, free Hamiltonians, $H(0, \Omega_{\mathrm{b}}, 0, 0)$ and 
$H(\Omega_{\mathrm{a}}, \Omega_{\mathrm{b}}, 0, 0)$, are defined by
\begin{align*}
  H(0, \Omega_{\mathrm{b}}, 0, 0)
  &= 1\otimes \hbar\Omega_{\mathrm{b}}\left( a^{\dagger}a+\frac{1}{2}\right), \\
  H(\Omega_{\mathrm{a}}, \Omega_{\mathrm{b}}, 0, 0)
  &=\frac{\hbar\Omega_{\mathrm{a}}}{2}\sigma_{z}\otimes 1+H(0, \Omega_{\mathrm{b}}, 0, 0),
\end{align*}
where $\Omega_{\mathrm{b}}$ denotes the frequency of the $1$-mode boson,
and $\Omega_{\mathrm{a}}$ is the atom transition frequency. 

It is easy to check that, for a common constant $\Omega>0$,
the Hamiltonian $H(\Omega, \Omega, 0, 0)$
has the $\mathcal{N}=2$\, SUSY, and the Hamiltonian $H(0, \Omega, 0, 0)$
makes its spontaneous breaking \cite{hir11,hir15}.
Here, the 2-level-system approximation
of the double-well potential works
for the spontaneous SUSY breaking,
which is explained in Section \ref{subsec:mechanism}. 
The individual algebraic structures are given in the following.

For the Hamiltonian $H(\Omega, \Omega, 0, 0)$,
its real supercharges, $q_{1}$ and $q_{2}$, are given by
$$
q_{1} =\sqrt{\frac{\hbar\Omega}{2}}\left(\sigma_{+}a+\sigma_{-}a^{\dagger}\right),\qquad
q_{2} =i\sqrt{\frac{\hbar\Omega}{2}}\left(\sigma_{-}a^{\dagger}-\sigma_{+}a\right).
$$
Then, they satisfy
\begin{align*}
  & \left\{ q_{k}, q_{\ell}\right\}
  =\delta_{k\ell}H(\Omega, \Omega, 0, 0), \\
  & \left[ q_{k}, H(\Omega, \Omega, 0, 0)\right]=0, \\
  & \left\{ q_{k}, N_{\mbox{\tiny F}}\right\}=0,
\end{align*}
where $N_{\mbox{\tiny F}}$ is the grading operator
defined by $N_{\mbox{\tiny F}}=-\sigma_{z}$.
The ground state (i.e., vacuum) ${\mid\downarrow\rangle}\otimes{\mid\!\!0\rangle}$
of $H(\Omega, \Omega, 0, 0)$
is a bosonic state since $N_{\mbox{\tiny F}}{\mid\downarrow\rangle}\otimes{\mid\!\!0\rangle}
=\, {\mid\downarrow\rangle}\otimes{\mid\!\!0\rangle}$, and
it satisfies $q_{k}\mid\downarrow\rangle\otimes\mid\!\!0\rangle=0$, $k=1, 2$. 
The complex supercharges, $q^{+}$ and $q^{-}$,
are given by
$$
  q^{+}=\frac{1}{\sqrt{2}}\left( q_{1}+iq_{2}\right)
  =\sqrt{\hbar\Omega}\, \sigma_{+}a,\qquad
  q^{-}=\frac{1}{\sqrt{2}}\left( q_{1}-iq_{2}\right)
  =\sqrt{\hbar\Omega}\, \sigma_{-}a^{\dagger},
$$
such that
\begin{align*}
  & H(\Omega, \Omega, 0, 0)=\left\{ q^{+}, q^{-}\right\}, \\
  & \left\{q^{\pm}, q^{\pm}\right\}=0, \\
  & \left[ H(\Omega, \Omega, 0, 0), q^{\pm}\right]=0.
\end{align*}
These complex supercharges make the connection
between the bosonic and fermionic states:
\begin{align*}
  & q^{-}\mid\downarrow\rangle\otimes\mid\!\!n\rangle
  =q^{+}\mid\uparrow\rangle\otimes\mid\!\!n\rangle=0, \\
  & \mid\uparrow\rangle\otimes\mid\!\!n\rangle
  =\frac{1}{\sqrt{(n+1)\hbar\omega\,}}
  q^{+}\mid\downarrow\rangle\otimes\mid\!\!n+1\rangle, \\
  & \mid\downarrow\rangle\otimes\mid\!\!n+1\rangle
  =\frac{1}{\sqrt{(n+1)\hbar\omega\,}}
  q^{-}\mid\uparrow\rangle\otimes\mid\!\!n\rangle.
\end{align*}
We immediately have
$q^{\pm}{\mid\downarrow\rangle}\otimes{\mid\!\!0\rangle}=0$
for the vacuum ${\mid\downarrow\rangle}\otimes{\mid\!\!0\rangle}$.
Since this vacuum is a unique ground state of
the Hamiltonian $H(\Omega, \Omega, 0, 0)$,
the Witten index is $1$.

\begin{figure}[ht]%
\centering
\includegraphics[width=0.5\textwidth]{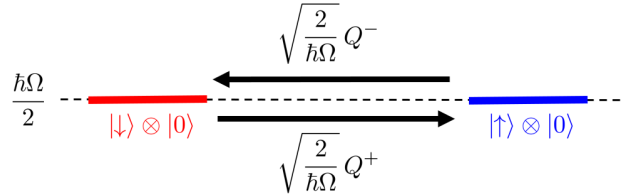}
\caption{Possibility of Existence of Goldstino.
  The excitation by supercharges $Q^{\pm}$ on vacuums,
  $\mid\downarrow\rangle\otimes\mid\!\!0\rangle$
  and $\mid\uparrow\rangle\otimes\mid\!\!0\rangle$,
  might make the Goldstino mode.}
\label{fig:goldstino_1}
\end{figure}

Meanwhile, the algebraic structure for the SUSY breaking of
$H(0, \Omega, 0, 0)$ is determined as follows:
Its real supercharges, $Q_{1}$ and $Q_{2}$, are given by
$$
  Q_{1}=\sqrt{\frac{\hbar\Omega}{2}}\,\sigma_{x}
  \sqrt{a^{\dagger}a+\frac{1}{2}},\qquad
  Q_{2}=\sqrt{\frac{\hbar\Omega}{2}}\,\sigma_{y}
  \sqrt{a^{\dagger}a+\frac{1}{2}}.
$$
Then, they satisfy
\begin{align*}
  & \left\{ Q_{k}, Q_{\ell}\right\}
  =\delta_{k\ell}H(0, \Omega, 0, 0), \\
  & \left[ Q_{k}, H(0, \Omega, 0, 0)\right]=0, \\
  & \left\{ Q_{k}, N_{\mbox{\tiny F}}\right\}=0,
\end{align*}
where $N_{\mbox{\tiny F}}$ is the grading operator defined by
$N_{\mbox{\tiny F}}=-\sigma_{z}$. 
The ground sates (i.e., vacuums) ${\mid\!\!\sharp\rangle}\otimes{\mid\!\!0\rangle}$,
$\sharp=\downarrow, \uparrow$,
of $H(0, \Omega, 0, 0)$ have the strictly positive, lowest eigenvalue
$\hbar\Omega/2>0$.
We have $Q_{k}{\mid\!\!\sharp\rangle}\otimes{\mid\!\!0\rangle}\ne 0$,
$k=1, 2$. 
The complex supercharges, $Q^{+}$ and $Q^{-}$,
are given by 
$$
Q^{\pm}=\frac{1}{\sqrt{2}}\left(Q_{1}\pm iQ_{2}\right)
=\sigma_{\pm}\sqrt{\hbar\Omega
  \left( a^{\dagger}a+\frac{1}{2}\right)}
$$
such that
\begin{align*}
  & H(0, \Omega, 0, 0)=\left\{ Q^{+}, Q^{-}\right\}, \\
  & \left\{Q^{\pm}, Q^{\pm}\right\}=0, \\
  & \left[ H(0, \Omega, 0, 0), Q^{\pm}\right]=0.
\end{align*}
These complex supercharges have the relations,   
$
Q^{-}\mid\downarrow\rangle\otimes\mid\!\!n\rangle
=Q^{+}{\mid\uparrow\rangle}\otimes\mid\!\!n\rangle=0
$.  
They do cut the connection with the boson annihilation and creation
but the connection between the bosonic and fermionic states as 
\begin{align*}
  & \mid\uparrow\rangle\otimes\mid\!\!n\rangle
  =\frac{1}{\sqrt{(n+\frac{1}{2})\hbar\Omega\,}}
  Q^{+}\mid\downarrow\rangle\otimes\mid\!\!n\rangle, \\
  & \mid\downarrow\rangle\otimes\mid\!\!n\rangle
  =\frac{1}{\sqrt{(n+\frac{1}{2})\hbar\Omega\,}}
  Q^{-}\mid\uparrow\rangle\otimes\mid\!\!n\rangle,
\end{align*}
in particular, $Q^{+}\mid\downarrow\rangle\otimes\mid\!\!0\rangle\ne 0$ and
$Q^{-}\mid\uparrow\rangle\otimes\mid\!\!0\rangle\ne 0$
for the vacuums
$\mid\!\!\sharp\rangle\otimes\mid\!\!0\rangle$,
$\sharp=\downarrow, \uparrow$. 
In terms of the grading operator $N_{\mbox{\tiny F}}$,
since $N_{\mbox{\tiny F}}{\mid\downarrow\rangle}\otimes{\mid\!\!n\rangle}=
{\mid\downarrow\rangle}\otimes{\mid\!\!n\rangle}$
and $N_{\mbox{\tiny F}}{\mid\uparrow\rangle}\otimes{\mid\!\!n\rangle}=
-{\mid\uparrow\rangle}\otimes{\mid\!\!n\rangle}$, 
the vacuum ${\mid\downarrow\rangle}\otimes{\mid\!\!0\rangle}$
is a bosonic state and
the vacuum ${\mid\uparrow\rangle}\otimes{\mid\!\!0\rangle}$
is a fermionic state. 
Thus, the Witten index is $0$,
and the SUSY is spontaneously broken.
The collaboration by the supercharges, $Q^{\pm}$,
can make the oscillation
between the degenerate ground states,
$\mid\downarrow\rangle\otimes\mid\!\!0\rangle$ and 
$\mid\uparrow\rangle\otimes\mid\!\!0\rangle$,
of the Hamiltonian $H(0, \Omega, 0, 0)$, 
which may emerge a fingerprint of
the Goldstino mode \cite{sal74, wit82, bin06, bau15a, bau15b, san16, san17, bla17, ma21, taj21}.
Briefly, since $H(0, \Omega, 0, 0){Q^{+}\!\!\mid\downarrow\rangle\otimes\mid\!\!0\rangle}
=(\hbar\Omega/2){Q^{+}\!\!\mid\downarrow\rangle\otimes\mid\!\!0\rangle}$
and 
$H(0, \Omega, 0, 0){Q^{-}\!\!\mid\uparrow\rangle\otimes\mid\!\!0\rangle}
=(\hbar\Omega/2){Q^{-}\!\!\mid\uparrow\rangle\otimes\mid\!\!0\rangle}$,
the states, ${Q^{+}\!\!\mid\downarrow\rangle\otimes\mid\!\!0\rangle}$
and ${Q^{-}\!\!\mid\uparrow\rangle\otimes\mid\!\!0\rangle}$,
are made up of the excitation of proper particles on
the vacuums, $\mid\downarrow\rangle\otimes\mid\!\!0\rangle$
and $\mid\uparrow\rangle\otimes\mid\!\!0\rangle$, respectively.
Since there is no energy increment between the individual vacuum and
the corresponding excited state by the supercharges $Q^{\pm}$, 
the particles might be Goldstinos (Fig.\ref{fig:goldstino_1}). 

\begin{figure}[ht]%
\centering
\includegraphics[width=0.55\textwidth]{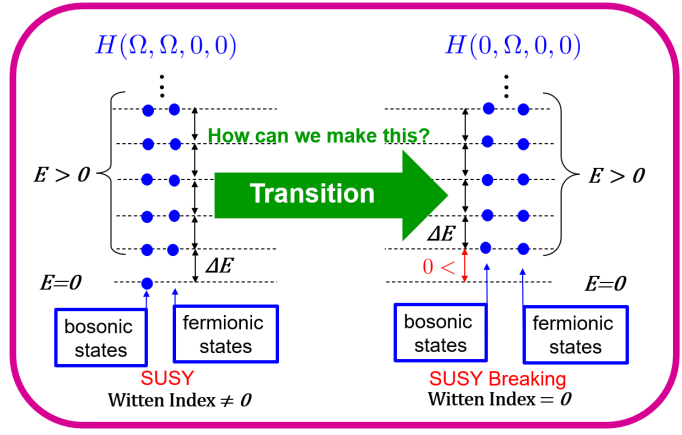}
\caption{Transition from SUSY to Its Spontaneous Breaking.
  The left schematic picture shows the energy-spectrum property of SUSY.
  The energy spectrum in the right schematic picture is for the spontaneous
  SUSY breaking.}
\label{fig:transition}
\end{figure}

Our problems are described in the following. \\
\textbf{Problem 1.} How can we introduce an interaction $H_{\mathrm{int}}$
between the $2$-level atom and $1$-mode boson
to make the transition (Fig.\ref{fig:transition}) from
the $\mathcal{N}=2$\, SUSY Hamiltonian $H(\Omega, \Omega, 0, 0)$ to
its spontaneous-breaking Hamiltonian unitarily equivalent to
the Hamiltonian $H(0, \Omega, 0, 0)$? \\
\textbf{Problem 2.} How can we make a mass term in the interaction
$H_{\mathrm{int}}$ which causes the mass enhancement in the SUSY breaking?

The prototype model in \cite{hir11,hir15} is proposed 
for a partial solution to Problem 1.
In this paper, thus, we extend it such that the extended model gives
a solution to Problems 1 {\&} 2. 

Our model is based on the quantum Rabi model
whose Hamiltonian is given by
$$
H_{\mbox{\tiny Rabi}}(\Omega_{\mathrm{a}}, \Omega_{\mathrm{b}}, G)
=H(\Omega_{\mathrm{a}}, \Omega_{\mathrm{b}}, 0, 0)
+\hbar G\sigma_{x}\left( a+a^{\dagger}\right),
$$
where the last term is the linear interaction between the atom and boson
with the parameter $G$ representing the coupling strength.
For our candidate of the interaction $H_{\mathrm{int}}$,
we add the quadratic interaction in addition to the linear one,
and thus, our total Hamiltonian reads
\begin{align}
  H(\Omega_{\mathrm{a}}, \Omega_{\mathrm{b}}, G, C)
  =& H_{\mbox{\tiny Rabi}}(\Omega_{\mathrm{a}}, \Omega_{\mathrm{b}}, G)
  +\hbar CG^{2}\left( a+a^{\dagger}\right)^{2},
  \label{eq:hamiltonian_0}
\end{align}
where the last term of Eq.(\ref{eq:hamiltonian_0}) is the quadratic interaction
$\hbar C\{G\sigma_{x}(a+a^{\dagger})\}^{2}$ with the parameter $C$
which controls the dimension and volume of the quadratic interaction energy.
This quadratic term is often called `$A^{2}$-term' \cite{rza75, nat10}.

As explained above, tuning the parameters $\Omega_{\mathrm{a}}$ and $\Omega_{\mathrm{b}}$  
as $\Omega_{\mathrm{a}}=\Omega_{\mathrm{b}}=\omega$
for a positive, common constant $\omega$,
the Hamiltonian $H(\omega, \omega, 0, 0)$
has the $\mathcal{N}=2$\, SUSY. 
In our model, as the coupling strength $G$ gets stronger enough,
the $A^{2}$-term may appear, i.e., $C\ne 0$.
Then, similarly to the case of the superradiant phase transition \cite{dicke54,hl73},
a no-go theorem caused by the $A^{2}$-term \cite{rza75}
should be minded for our target transition.
In that case, its avoidance should be argued for our model described by Eq.(\ref{eq:hamiltonian_0}) in SUSY QM
as well as for the superradiant-phase-transition model \cite{nat10}.
We investigate this problem from now on.

As shown in Eq.(2) of Ref.\cite{hir22}, 
for every non-negative $C$,
we have a unitary operator $U_{A^{2}}$ such that
\begin{eqnarray}
  U_{A^{2}}^{*}H(\Omega_{\mathrm{a}}, \Omega_{\mathrm{b}}, G, C)U_{A^{2}}
  &=&H(\Omega_{\mathrm{a}}, \Omega(G), \widetilde{G}, 0)
  =H_{\mbox{\tiny Rabi}}(\Omega_{\mathrm{a}}, \Omega(G), \widetilde{G}),
  \label{eq:hb-trans}
\end{eqnarray}
where
$\Omega(G)=\sqrt{\Omega_{\mathrm{b}}^{2}+4C\Omega_{\mathrm{b}}G^{2}\,}$
and $\widetilde{G}=G\sqrt{\Omega_{\mathrm{b}}/\Omega(G)}$.
In the same way as in Eq.(3) of Ref.\cite{hir22}, 
for the displacement operator $D(G/\Omega_{\mathrm{b}}) =
\exp\left[ G(a^{\dagger}-a)/\Omega_{\mathrm{b}}\right]$,
we can define a unitary operator $U(G/\Omega_{\mathrm{b}})$ by
$$
U(G/\Omega_{\mathrm{b}})=\frac{1}{\sqrt{2}}
\left\{\left(\sigma_{-}-1\right)\sigma_{+}D(G/\Omega_{\mathrm{b}})
+\left(\sigma_{+}+1\right)\sigma_{-}D(-G/\Omega_{\mathrm{b}})\right\},
$$
and then, we obtain the equation, 
\begin{eqnarray}
  &{}&
  U(G/\Omega_{\mathrm{b}})^{*}
  \left\{
  H(\Omega_{\mathrm{a}}, \Omega_{\mathrm{b}}, G, 0)
  +\hbar\frac{G^{2}}{\Omega_{\mathrm{b}}}
  \right\}
  U(G/\Omega_{\mathrm{b}})
  \nonumber \\ 
  &=&
  H(0, \Omega_{\mathrm{b}}, 0, 0)
    -\frac{\hbar\Omega_{\mathrm{a}}}{2}
  \left\{\sigma_{+}D(G/\Omega_{\mathrm{b}})^{2}
  +\sigma_{-}D(-G/\Omega_{\mathrm{b}})^{2}\right\}.
  \label{eq:ut}
\end{eqnarray}

From now on,
following Ref.\cite{hir22},
we will explain the no-go theorem and its avoidance.

\subsection{No-go theorem in strong coupling limit}

Now we consider the strong coupling limit for
the quantum Rabi model without and with the $A^{2}$-term.
This limit is approximately realized in experiments of the deep-strong coupling regime \cite{cas10},
for instance, in circuit QED \cite{yos17}. 
The parameters, $\Omega_{\mathrm{a}}$, $\Omega_{\mathrm{b}}$, and $G$, are set
as $\Omega_{\mathrm{a}}=\Omega_{\mathrm{b}}=\omega$
and $G=\mathrm{g}$ for a non-negative parameter $\mathrm{g}$.
The Hamiltonian $H(\omega, \omega, \mathrm{g}, 0)
=H_{\mbox{\tiny Rabi}}(\omega, \omega, \mathrm{g})$ is for
the quantum Rabi model, and denoted by
$H_{\mbox{\tiny Rabi}}(\mathrm{g})$ for simplicity.
In the renormalization for the $A^{2}$-term,
the quantities $\Omega(\mathrm{g})$ and
$\widetilde{\mathrm{g}}$ are defined by $\Omega(\mathrm{g})=
\sqrt{\omega^{2}+4C\omega\mathrm{g}^{2}\,}$ and 
$\widetilde{\mathrm{g}}=\mathrm{g}\sqrt{\omega/\Omega(\mathrm{g})}$.

\begin{figure}[ht]%
\centering
\includegraphics[width=0.45\textwidth]{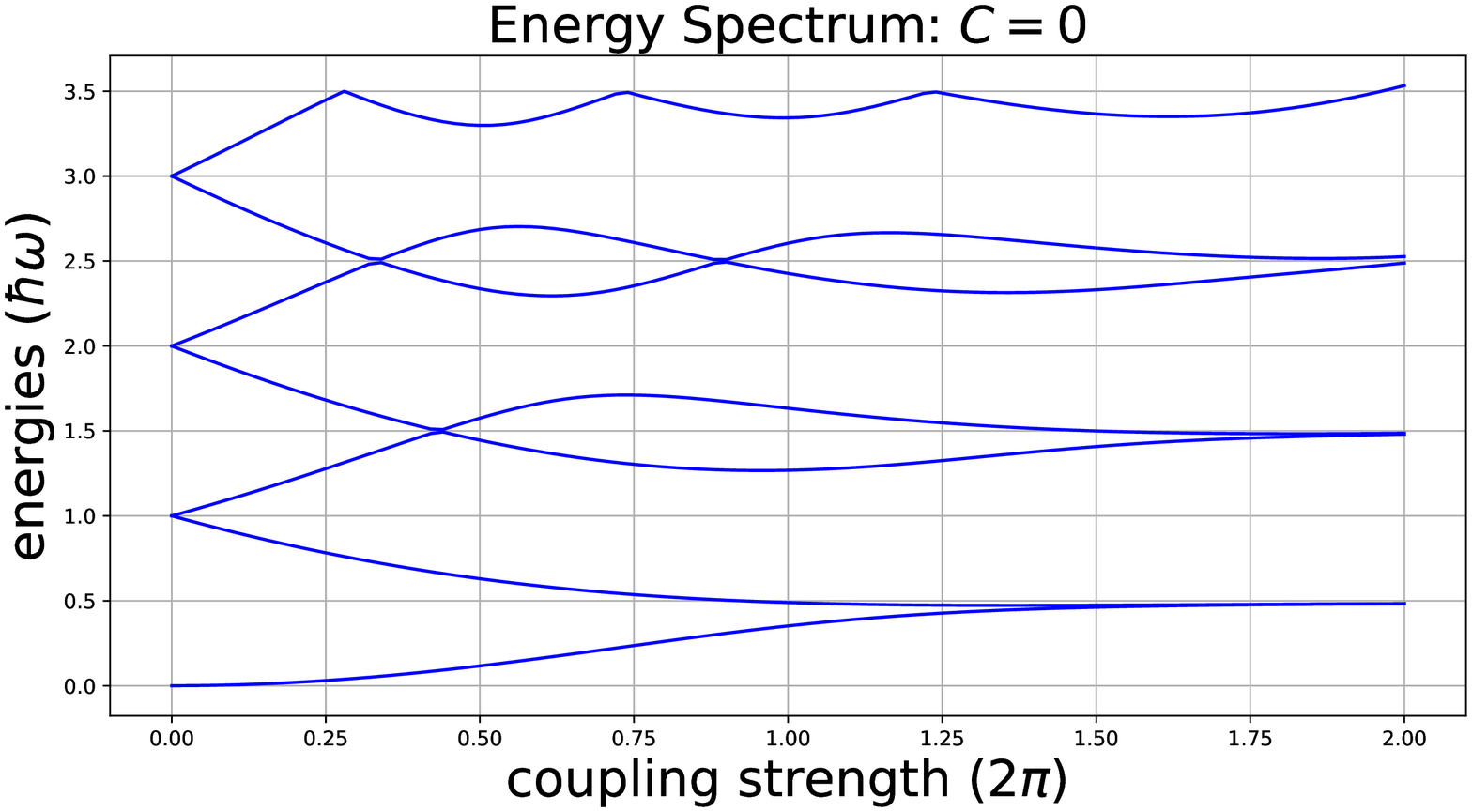}
\quad
\includegraphics[width=0.45\textwidth]{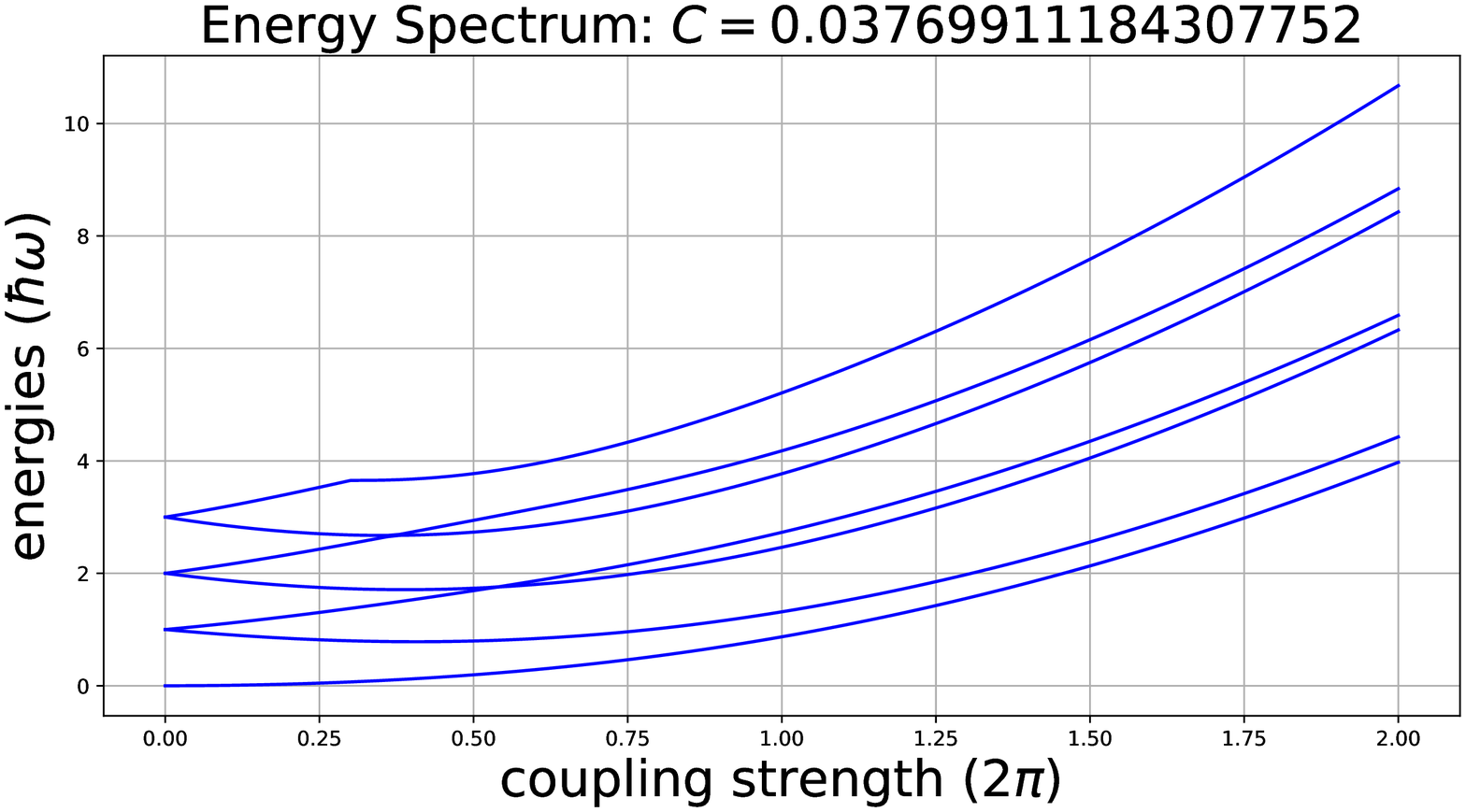}
\vspace*{17mm}
\caption{Energy Spectrum of
    $H_{\mbox{\tiny Rabi}}(\mathrm{g})+\hbar C\mathrm{g}^{2}
    \left( a+a^{\dagger}\right)^{2}+\hbar\widetilde{\mathrm{g}}^{2}/\Omega(\mathrm{g})$ with
    $\omega=6.2832$. 
    A ground state energy and six excited state energies
    from the bottom are shown in each graph. 
The left graph shows the energy spectrum for $C=0$. 
The right graph is for $C=0.0377$.
The left graph says that the quantum Rabi model (without $A^{2}$-term)
has the transition from the $\mathcal{N}=2$\, SUSY to its spontaneous breaking.
On the other hand, the right graph shows the loss of the spontaneous breaking, 
and reveals the explicit breaking instead.
Here, it should be noted $\lim_{\mathrm{g}\to\infty}\hbar\Omega(\mathrm{g})=\infty$ 
and $\lim_{\mathrm{g}\to\infty}\hbar\widetilde{\mathrm{g}}^{2}/\Omega(\mathrm{g})
=\hbar/(4C)$.}\label{fig:Fig_g}
\end{figure}

In case $C=0$, the quantum Rabi Hamiltonian with its self-energy,
$H_{\mbox{\tiny Rabi}}(\mathrm{g})+\hbar\frac{\mathrm{g}^{2}}{\omega}$,
is asymptotically equal to the Hamiltonian, 
$U(\mathrm{g}/\omega)H(0, \omega, 0, 0)U(\mathrm{g}/\omega)^{*}$,
as $\mathrm{g}\to\infty$.
Thus, the $\mathcal{N}=2$\, SUSY is spontaneously broken
in the strong coupling limit $\mathrm{g}\to\infty$.
This is completely characterized with the energy-spectrum property,
for instance, as in the left graph of Fig.\ref{fig:Fig_g}. 
We here note that the ground states of
$U(\mathrm{g}/\omega)H(0, \omega, 0, 0)U(\mathrm{g}/\omega)^{*}$
become the Schr\"{o}dinger-cat-like states.

In the case $C>0$, on the other hand,
the quantum Rabi Hamiltonian with its self-energy and the $A^{2}$-term,
$H_{\mbox{\tiny Rabi}}(\mathrm{g})
+\hbar C\mathrm{g}^{2}\left( a+a^{\dagger}\right)^{2}
+\hbar\frac{\widetilde{\mathrm{g}}^{2}}{\Omega(\mathrm{g})}$,
is asymptotically equal to the Hamiltonian,
$U_{A^{2}}
U(\widetilde{\mathrm{g}}/\Omega(\mathrm{g}))
\left[
H(0, \Omega(\mathrm{g}), 0, 0)
-\, \frac{\hbar\omega}{2}\sigma_{x}
\right]
U(\widetilde{\mathrm{g}}/\Omega(\mathrm{g}))^{*}
U_{A^{2}}^{*}$,
as $\mathrm{g}\to\infty$.
The appearance of the atomic term $\hbar\omega\sigma_{x}/2$ interferes with
the transition to the SUSY breaking.
Moreover, the divergence of $\Omega(\mathrm{g})$ as $\mathrm{g}\to\infty$,
together with the atomic term,
rudely crushes and explicitly breaks that SUSY.
We can see this crush in the energy spectrum, for instance,
as in the right graph of Fig.\ref{fig:Fig_g}.
Thus, the above quantum Rabi model with 
the $A^{2}$-term cannot go to
the SUSY breaking as $\mathrm{g}$
changes from $\mathrm{g}=0$ to $\mathrm{g}\approx\infty$.
This is the `no-go theorem'  caused by the $A^{2}$-term
for the SUSY breaking in the strong coupling limit.

These results are mathematically established using the limit in the norm resolvent sense,
and the limit is valid over the energy spectrum (see Theorem VIII.24 of \cite{rs1}). 
Thus, the limit energy spectra are obtained by the individual, asymptotic equalities.
Whether the $\mathcal{N}=2$\, SUSY of $H(\omega, \omega, 0, 0)$ is taken
to its spontaneous breaking is checked by seeing the energy degeneracy
and measuring each interval between adjacent energy levels. 
The energy spectrum by the numerical computations
with QuTiP \cite{nori1, nori2} is obtained, for instance,
as in Fig.\ref{fig:Fig_g}.

\subsection{Limit for avoidance of no-go theorem}

In order to avoid the no-go theorem, as shown in Ref.\cite{hir22},
we employ the limit used in Ref.\cite{cai22} 
experimentally to realize the transition
for the prototype model.
We prepare a continuous function $\omega[r]$
of $1$-variable $r$, $0\le r\le 1$, such that $\omega[0]=\omega$
and $\omega[1]=0$.
Then, the Hamiltonian $H(\omega[0], \omega, 0, 0)
=H(\omega, \omega, 0, 0)$ has the $\mathcal{N}=2$\, SUSY,
and the Hamiltonian $H(\omega[1], \omega, 0, 0)
=H(0, \omega, 0, 0)$ makes its spontaneous breaking.
Cai \textit{et al.} have the trapped-ion technology
to realize this limit in the case $C=0$ \cite{cai22}. 
Indeed the linear interaction cannot, alone, do anything to enhance the mass, 
but it works for the mass enhancement not only in the bosonic states but also
in the fermionic states with the help of the $A^{2}$-term.
We explain this in Section \ref{subsec:mechanism}.

Let $g(r)$ be a continuous function of $1$-variable $r$,
$0\le r \le 1$, satisfying $g(0)=0$ and $g(1)=\mathrm{g}$. 
The parameters, $\Omega_{\mathrm{a}}$, $\Omega_{\mathrm{b}}$, and $G$, 
are given by $\Omega_{\mathrm{a}}=\omega[r]$, $\Omega_{\mathrm{b}}=\omega$,
and $G=g(r)$.
The Hamiltonian $H(\omega[r], \omega, g(r), 0)=H_{\mbox{\tiny Rabi}}(\omega[r], \omega, g(r))$
for the quantum Rabi model 
is denoted by $H_{\mbox{\tiny Rabi}}[r]$ for simplicity.
The renormalized quantities $\widetilde{\omega}[r]$ and
$\widetilde{g}[r]$ are given by 
$\widetilde{\omega}[r]=
\sqrt{\omega^{2}+4C\omega g(r)^{2}\,}$
and   
$\widetilde{g}[r]=g(r)\sqrt{\omega/\widetilde{\omega}[r]}$.
Then, Eq.(6) of Ref.\cite{hir22} says    
\begin{eqnarray}
  &{}&
  H_{\mbox{\tiny Rabi}}[r]
  +\hbar Cg(r)^{2}\left( a+a^{\dagger}\right)^{2}
  +\hbar\frac{\widetilde{g}[r]^{2}}{\widetilde{\omega}[r]}
  \nonumber \\ 
  &\longrightarrow&
  U_{A^{2}}
  U(\widetilde{\mathrm{g}}[1]/\widetilde{\omega}[1])
  H(0, \widetilde{\omega}[1], 0, 0)
  U(\widetilde{\mathrm{g}}[1]/\widetilde{\omega}[1])^{*}
  U_{A^{2}}^{*}  
  \label{eq:approx3}
\end{eqnarray}
in the norm resolvent sense \cite{rs1} as $r\to 1$.
It is worthy to note that the ground states of
$U_{A^{2}}
  U(\widetilde{\mathrm{g}}[1]/\widetilde{\omega}[1])
  H(0, \widetilde{\omega}[1], 0, 0)
  U(\widetilde{\mathrm{g}}[1]/\widetilde{\omega}[1])^{*}
  U_{A^{2}}^{*}$
are obtained as the unitary transformation
of Schr\"{o}dinger-cat-like states.

\begin{figure}[ht]%
\centering
\includegraphics[width=0.45\textwidth]{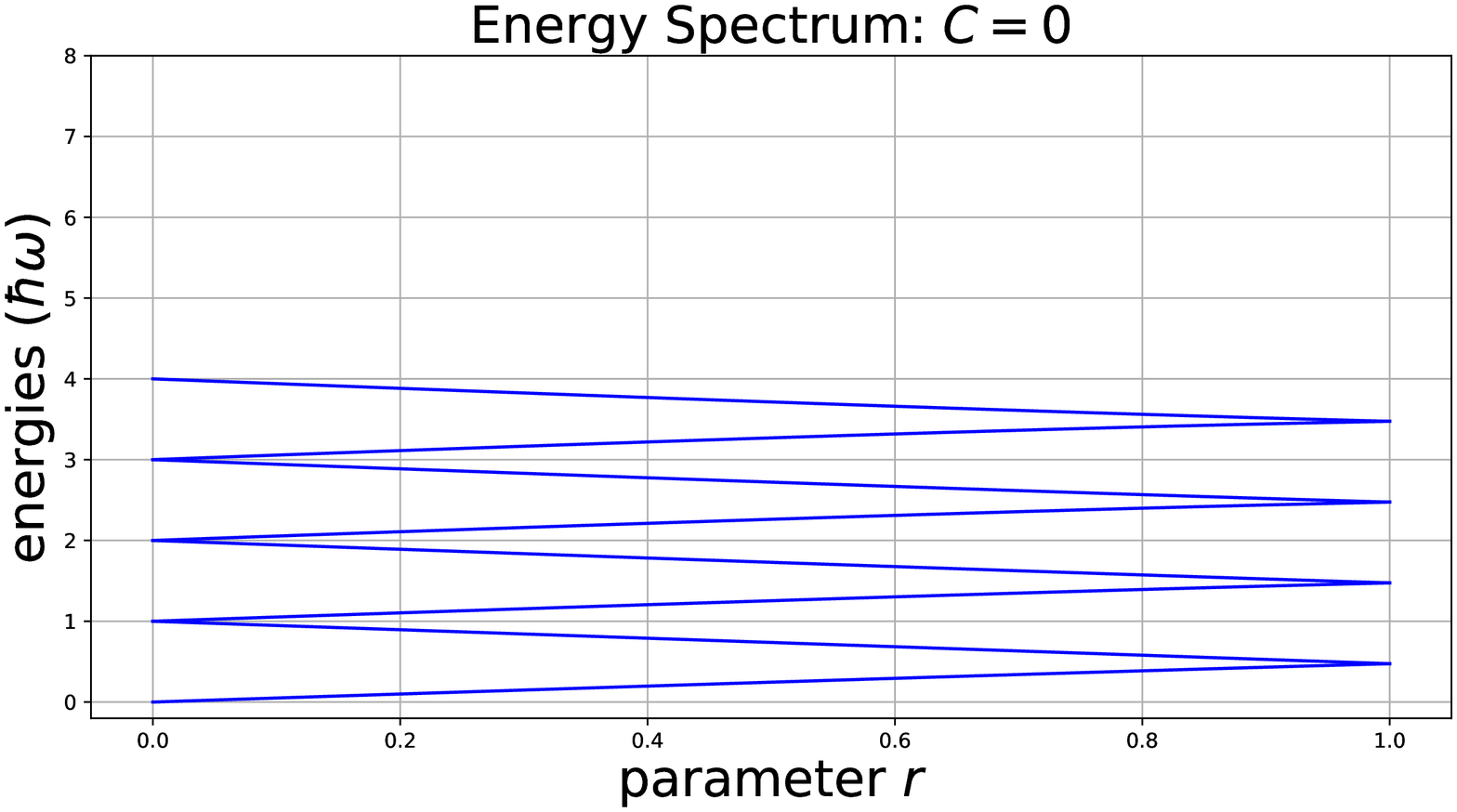}
\quad
\includegraphics[width=0.45\textwidth]{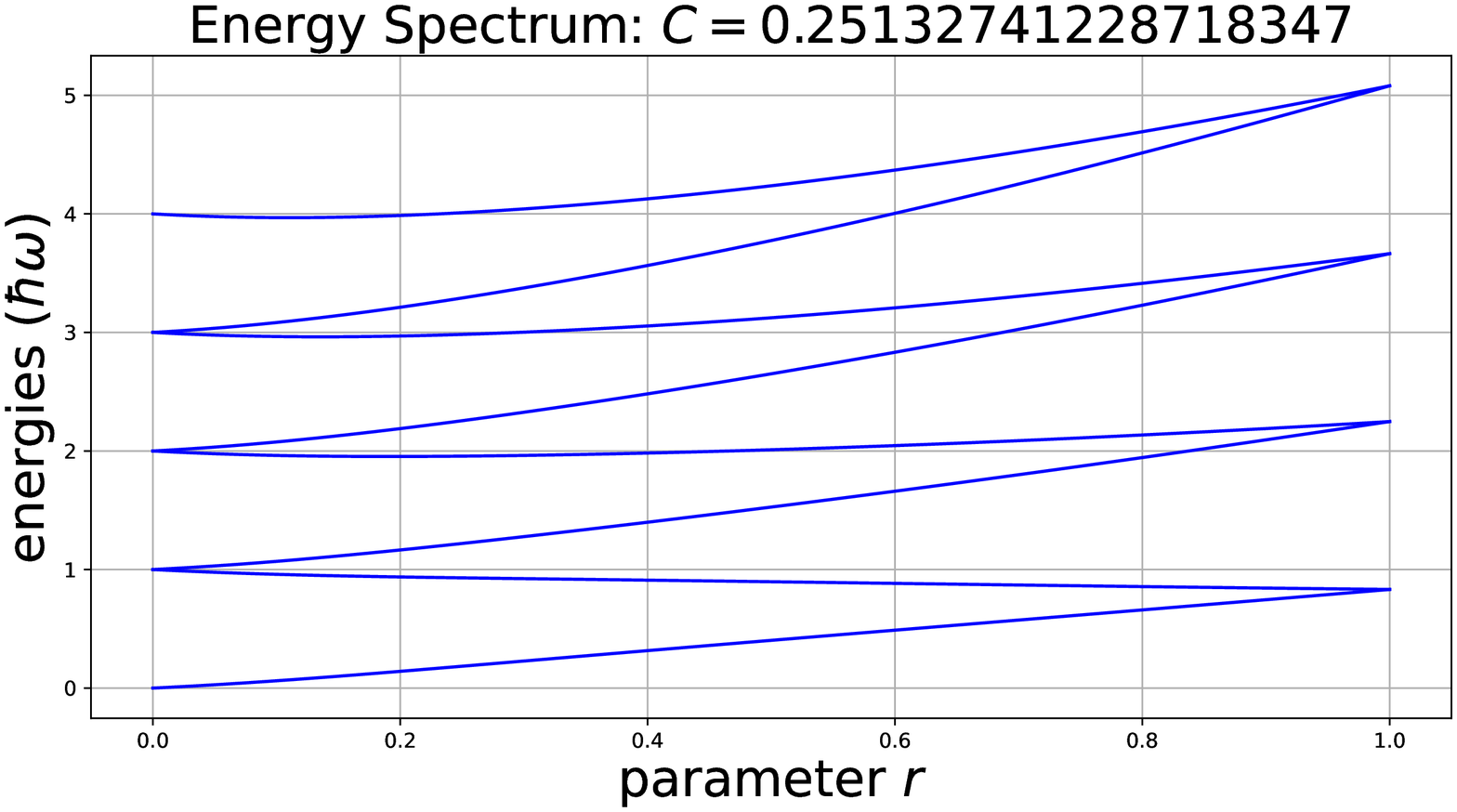}
\vspace*{17mm}
\caption{Energy Spectrum of
    $H_{\mbox{\tiny Rabi}}[r]+\hbar Cg(r)^{2}
  \left( a+a^{\dagger}\right)^{2}+\hbar\widetilde{g}[r]^{2}/\widetilde{\omega}[r]$ with $\omega=6.2832$ and $\mathrm{g}=6.2832$. 
  A ground state energy and six excited state energies
    from the bottom are shown in each graph.
The left graph shows the energy spectrum for $C=0$. 
The right graph is for $C=0.2513$.
The quantum Rabi models without $A^{2}$-term (i.e., $C=0$)
and with $A^{2}$-term (i.e., $C>0$)  
have the transition from the $\mathcal{N}=2$\, SUSY to its spontaneous breaking.
In particular, the energy gap by the $A^{2}$-term
appears in $\hbar\widetilde{\omega}[1]$ of the right graph.
In these numerical computations,
we employ $\omega[r]=(1-r)\omega$
and $g(r)=r\mathrm{g}$.}\label{fig:Fig_r}
\end{figure}

Eq.(\ref{eq:approx3}) says that the Hamiltonian $H(0, \widetilde{\omega}[1], 0, 0)$
appears in the limit,
and therefore, 
the Rabi model with $A^{2}$-term, described by
$H_{\mbox{\tiny Rabi}}[r]
+\hbar Cg(r)^{2}\left( a+a^{\dagger}\right)^{2}
+\hbar\widetilde{g}[r]^{2}/\widetilde{\omega}[r]$,
yields the SUSY breaking in the limit $r\to 1$.
The limit in the norm resolvent sense guarantees the convergence
of each energy level (see Theorem VIII.24 of \cite{rs1}). 
Thus, it is worthy to note that how the transition
from the $\mathcal{N}=2$\, SUSY to its spontaneous breaking
takes place,
and how the energy gap is produced in that transition.
The energy gap is governed by the parameter $C$ of
the $A^{2}$-term.
The energy spectrum is checked with QuTiP \cite{nori1, nori2},
for instance, as in Fig.\ref{fig:Fig_r}.
In particular, the comparison of the two graphs of Fig.\ref{fig:Fig_r}
shows the energy gap caused by the $A^{2}$-term.

\section{Radiative mass-enhancement in SUSY breaking}\label{sec3}

\subsection{Mathematical model for quantum simulation}

We consider the position operator $X$ and the momentum operator $P$ 
acting in the boson Fock space $\mathcal{F}_{\mathrm{b}}$,
and identify them with $1\otimes X$ and $1\otimes P$ acting in the state space
$\mathbb{C}^{2}\otimes\mathcal{F}_{\mathrm{b}}$, respectively.
For these identified position and momentum operators, $X$ and $P$, 
we give the Hamiltonian $H$ of a harmonic oscillator.
This describes the energy operator of a $1$-mode massive boson.
It is given by 
\begin{equation}
  H
  = \left(\frac{1}{2}P^{2}+\frac{\omega_{\mathrm{g}}^{2}}{2}X^{2}\right)
  \label{eq:h_XP}
\end{equation}
acting in the state space $\mathbb{C}^{2}\otimes\mathcal{F}_{\mathrm{b}}$, 
where $\hbar\omega_{\mathrm{g}}$ is the boson energy.
We call this $1$-mode massive boson the `heavy boson.'

We arbitrarily give a positive parameter $\omega$,
a non-negative parameter $C$, and a positive constant
$\mathrm{g}$ such that
$\omega_{\mathrm{g}}^{2}=\omega^{2}+4C\omega\mathrm{g}^{2}$. 
We consider another Hamiltonian $H_{\mbox{\tiny SS}}$
for the position operator $x$ and the momentum operator $p$ acting in
another boson Fock space $\mathcal{F}_{\mathrm{b}}$.
The Hamiltonian $H_{\mbox{\tiny SS}}$ is
popular in SUSY QM \cite{bin06, gmr11}
and given by 
\begin{equation}
H_{\mbox{\tiny SS}}
=1\otimes\frac{1}{2}\left( p^{2}+W^{2}\right)
+\frac{\hbar}{2}\sigma_{z}\otimes\frac{d W}{dx}, 
\label{eq:h_ss}
\end{equation}
where $W$ is the superpotential given by $W(x)=\omega x$.
We omit `$\otimes$', and then,
$H_{\mbox{\tiny SS}}=(1/2)\left(
p^{2}+W^{2}+\hbar\sigma_{z}(dW/dx)\right)$. 

Our spin-boson interaction is based on $\sigma_{x}x$.
It should be pointed out that the Pauli matrix $\sigma_{x}$
plays a role of the swap between the bosonic and fermionic states. 
We suppose that an extra second-order term $(2C/\omega)g(r)^{2}(\sigma_{x}W)^{2}
=(2C/\omega)g(r)^{2}W^{2}$, 
different from the second-order term $W^{2}/2$ in Eq.(\ref{eq:h_ss}), 
appears in our interaction as well as the first-order term
$g(r)\sqrt{2\hbar/\omega}\, \sigma_{x}W$. 
We prepare an interaction,
\begin{equation}
H_{\mathrm{int}}(r)
=g(r)\sqrt{\frac{2\hbar}{\omega}}\, \sigma_{x}W
+\frac{2C}{\omega}g(r)^{2}W^{2}
+\frac{\hbar g(r)^{2}}{4Cg(r)^{2}+\omega}
+\frac{\hbar}{2}\sigma_{z}\frac{dW_{\mathrm{a}}(r)}{dx},
\label{eq:h_int}
\end{equation}
for $r$, $0\le r\le 1$, with functions of $r$, $g(r)$,
$W_{\mathrm{a}}(r)=\left(\omega_{\mathrm{a}}(r)-\omega\right)x$,
and $\omega_{\mathrm{a}}(r)$. 
This interaction $H_{\mathrm{int}}(r)$ is introduced to cause a SUSY breaking
for the SUSY Hamiltonian $H_{\mbox{\tiny SS}}$.
Unlike Nambu and Jona-Lasinio's case \cite{nam61} and Goldstone's \cite{gol61},
the interaction $H_{\mathrm{int}}(r)$ has no Higgs potential.
However, as explained in Section \ref{subsec:mechanism},
it describes the boson interacting to the qubit (i.e., 2-level system),
and thus, the square of the superpotential $W^{2}$ acting in
$\mathbb{C}^{2}\otimes \mathcal{F}_{\mathrm{b}}$ 
makes the 2-level-system approximation of the double-well potential
(Fig.\ref{fig:approx_Higgs}).
We actually need a change of both the well shape for our potential
because the 2-level-system approximation
is just an approximation, not a true Higgs potential. 
In terms of oscillator, the second-order term, $(2C/\omega)g(r)^{2}W^{2}$,
means that the oscillator is coupled not only to its nearest neighbor but also
to itself at the equilibrium points, and induces a mass (see Chapter 3 of \cite{ht62}).  
Thus, we expect the extra second-order term in $H_{\mathrm{int}}(r)$ to play a role of
radiatively making the mass enhancement.
Our total Hamiltonian reads
$$
H(r)=H_{\mbox{\tiny SS}}+H_{\mathrm{int}}(r)
$$
then.
We control the interaction appearance using the functions $g(r)$ and $\omega_{a}(r)$.
Here, $g(r)$ is a continuous function satisfying
$g(0)=0$ and $g(1)=\mathrm{g}$.
The function $\omega_{\mathrm{a}}(r)$ is also continuous and satisfies 
$\omega_{\mathrm{a}}(0)=\omega$ and $\omega_{\mathrm{a}}(1)=0$.
Then, the total Hamiltonian attains the SUSY Hamiltonian at $r=0$: $H(0)=H_{\mbox{\tiny SS}}$.

We bring up the parameter $r$ from $r=0$ to $r=1$
in the total Hamiltonian $H(r)$. 
Following the mathematical methods \cite{hir15, hir17, hir20}, 
we can show $H(r)\to H(1)$ as $r\to 1$
in the norm resolvent sense \cite{rs1}.
As shown below, actually, $H(1)=H$.
In the case $C=0$, it can mathematically be proved that
this limit produces the transition
from the $\mathcal{N}=2$\, SUSY at $r=0$
to its spontaneous breaking at $r=1$
in the same way as in \cite{hir15, cai22}.
Cai \textit{et al.} report its two kinds of experimental observations
in a trapped ion quantum simulator \cite{cai22}.
The condition $C=0$ means that there is no mass-enhancement tool
in the interaction $H_{\mathrm{int}}(r)$,
and there is no possibility that the SUSY breaking can
yields a mass enhancement.
In the case $C>0$, however, there is that possibility.
We check this possibility below. 
Thus, we allocate the mass-enhancement role to the second-order term,
$(2C/\omega)g(r)^{2}W^{2}$ with $C>0$, in our model,
and then, we will theoretically show that for $C>0$
the mass enhancement in the fermionic states as well 
as in the bosonic ones takes place
in the process of the transition from $\mathcal{N}=2$\, SUSY to its
spontaneous breaking.

We consider the limit,
$H(r)\to H(1)$ as $r\to 1$. 
Defining the $1$-mode boson annihilation operator $B$ by
$$
B=\sqrt{\frac{\omega_{\mathrm{g}}}{2\hbar}}\, X+i\sqrt{\frac{1}{2\hbar\omega_{\mathrm{g}}}}\, P,
$$
the Hamiltonian $H$ in Eq.(\ref{eq:h_XP}) of the heavy boson can be rewritten as
$$
H=\hbar\omega_{\mathrm{g}}\left( B^{\dagger}B+\frac{1}{2}\right).
$$
Meanwhile, we define the $1$-mode boson annihilation operator
$b$ by
$$
b=\sqrt{\frac{\omega}{2\hbar}}\, x+i\sqrt{\frac{1}{2\hbar\omega}}\, p.
$$
We call this $1$-mode boson the `light boson'
compared with the heavy boson. 
Then, we can rewrite the total Hamiltonian $H(r)$ of the light boson as
$$
H(r)
=H_{\mbox{\tiny Rabi}}(r)
+\hbar Cg(r)^{2}( b+b^{\dagger})^{2}
+\frac{\hbar g(r)^{2}}{4Cg(r)^{2}+\omega},
$$
where $H_{\mbox{\tiny Rabi}}(r)$ is the Hamiltonian of
the quantum Rabi model \cite{rab36, rab37, bra11} given by
$$
H_{\mbox{\tiny Rabi}}(r)=\hbar\omega\left( b^{\dagger}b+\frac{1}{2}\right)
+\hbar g(r)\sigma_{x}\left( b+b^{\dagger}\right)
+\frac{\hbar\omega_{\mathrm{a}}(r)}{2}\sigma_{z}.
$$
The total Hamiltonian $H(r)$ is unitarily equivalent to the Hamiltonian
$H(\omega_{\mathrm{a}}(r), \omega, g(r), C)+\hbar g(r)^{2}/(4Cg(r)^{2}+\omega)$,
where the definition of $H(\omega_{\mathrm{a}}(r), \omega, g(r), C)$
is given in Section \ref{sec2}.
The second-order term $\hbar Cg(r)^{2}( b+b^{\dagger})^{2}$
is the $A^{2}$-term \cite{rza75, nat10}.
The $A^{2}$-term naturally appears in quantum electrodynamics (QED) 
and cavity QED when the coupling strength $g(r)$ is not so small.
Moreover, it may be controlled in circuit QED (see \cite{nat10}
and Methods of \cite{yos17}).

For every $r$ with $0\le r\le 1$,
we prepare functions, $\omega_{\mathrm{g}}(r)$ and $\widetilde{g}(r)$,
of $1$-variable $r$ by 
$\omega_{\mathrm{g}}(r)=\sqrt{\omega^{2}+4C\omega g(r)^{2}}$ 
and $\widetilde{g}(r)=g(r)\sqrt{\omega/\omega_{\mathrm{g}}(r)}$.
Replacing $\Omega_{\mathrm{a}}$, $\Omega_{\mathrm{b}}$,
$G$, $\Omega(G)$, and $\widetilde{G}$ in Eqs.(\ref{eq:hb-trans}) and (\ref{eq:ut})
by $\omega_{\mathrm{a}}(r)$, $\omega$,
$g(r)$, $\omega_{\mathrm{g}}(r)$, and $\widetilde{g}(r)$, respectively, 
we can make the unitary operator $U_{r}$,
and define a boson annihilation operator $B_{r}$ and the spin operators
$\mathcal{D}_{\pm}$ by
\begin{align}
  B_{r}&=U_{r}bU_{r}^{*}=(c_{1}+c_{2})b+(c_{1}-c_{2})b^{\dagger}
  +\frac{\widetilde{g}(r)}{\omega_{\mathrm{g}}(r)}\, \sigma_{x}, 
  \label{eq:unitrans1a} \\
  \mathcal{D}_{\pm}&=U_{r}\sigma_{\pm}
  \exp\left[\pm 2\frac{\widetilde{g}(r)}{\omega_{\mathrm{g}}(r)}
    \left( b^{\dagger}-b\right)\right]U_{r}^{*}
  =-\, \frac{1}{2}\left(\sigma_{z}\mp i\sigma_{y}\right), 
  \label{eq:unitras1b}
\end{align}
where $c_{1}=(1/2)\sqrt{\omega_{\mathrm{g}}(r)/\omega}$ and 
$c_{2}=(1/2)\sqrt{\omega/\omega_{\mathrm{g}}(r)}$.
Then, $B_{1}$ is unitarily equivalent to $B$
since $\omega_{\mathrm{a}}(1)=0$, $g(1)=\mathrm{g}$,
$\omega_{\mathrm{g}}(1)=\omega_{\mathrm{g}}$,
and $\widetilde{g}(1)=\widetilde{\mathrm{g}}\equiv\mathrm{g}\sqrt{\omega/\omega_{\mathrm{g}}}$. 
Thus, we identify $B_{1}$ with $B$, i.e., $B_{1}=B$, from now on.

We note that the canonical commutation relation and 
canonical anticommutation relation respectively hold: 
\begin{align*}
  \left[ B_{r}, B_{r}^{\dagger}\right]=\left[ b, b^{\dagger}\right]=1,\qquad 
  \left\{ \mathcal{D}_{-}, \mathcal{D}_{+}\right\}=1,\qquad
  \left\{ \mathcal{D}_{\pm}, \mathcal{D}_{\pm}\right\}=0.
\end{align*}
In addition, we realize the spin-chiral symmetry,
$$
[\sigma_{x}, B_{r}]=[\sigma_{x}, B_{r}^{\dagger}]=0.
$$
In other words, it is the symmetry with respect to the swap
between the bosonic and fermionic states. 
Eq.(\ref{eq:unitrans1a}) says that the boson annihilation operator $B_{r}$
consists of the pair of the annihilation and creation of the light boson with the $X$-gate.
This pair is produced following the (meson) pair theory \cite{ht62, hir17, hir20}.
Since $B_{1}=B$ in particular, we can think that the heavy boson is a quasi-particle of
the annihilation and creation of the light bosons which eats $\sigma_{x}$. 
Eq.(\ref{eq:unitras1b}) says that the heavy boson cannot see the displacement by the light boson
directly in the spin.

Then, we have the equation between the Hamiltonian described by the light boson coupled with the spin
and the Hamiltonian described by the heavy boson coupled with the spin, 
\begin{align}
  \hbar\omega_{\mathrm{g}}(r)\left( B_{r}^{\dagger}B_{r}+\frac{1}{2}\right)
  -\, \frac{\hbar\omega_{\mathrm{a}}(r)}{2}
  \left(\mathcal{D}_{-}+\mathcal{D}_{+}\right)
  = H(r).
  \label{eq:unitrans2}
\end{align}

We have $\omega_{\mathrm{a}}(1)=0$, and $\omega_{\mathrm{g}}(1)=\omega_{\mathrm{g}}$
because $g(1)=\mathrm{g}$.
Thus, we obtain the limit
\begin{align}
H(r)&= H_{\mbox{\tiny Rabi}}(r)+\hbar Cg(r)^{2}( b+ b^{\dagger})^{2}
+\frac{\hbar g(r)^{2}}{4Cg(r)^{2}+\omega}
\nonumber \\ 
&\longrightarrow
H=\hbar\omega_{\mathrm{g}}\left( B^{\dagger}B+\frac{1}{2}\right)
\label{eq:important-limit}
\end{align}
as $r\to 1$.
This limit is consistent with Eq.(\ref{eq:approx3}) and its rephrasing
in the present case.

\subsection{Mechanism of radiative mass-enhancement}
\label{subsec:mechanism}

\begin{figure}[ht]%
\centering
\includegraphics[width=0.85\textwidth]{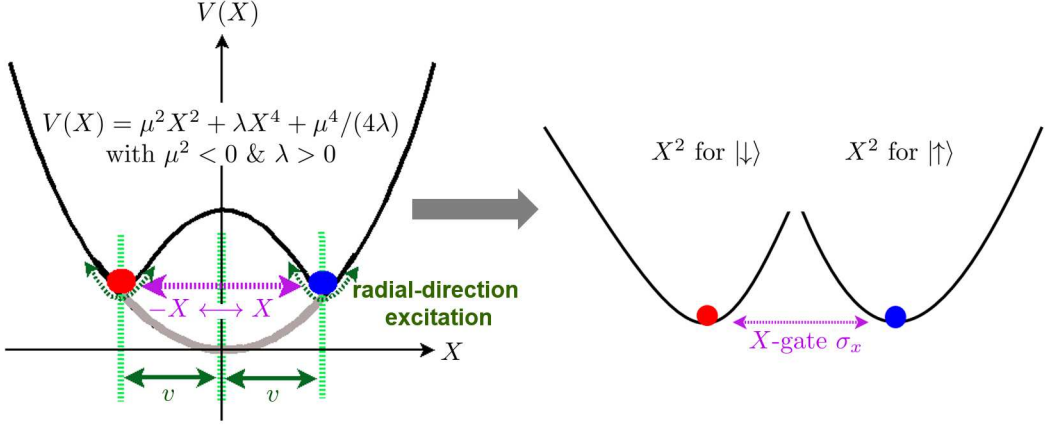}
\caption{2-Level-System Approximation.
  The left graph is the schematic image
  of the cross section of the Higgs potential 
  with the $XV$-plane.
  The right graph is the schematic notion of the 2-level-system
  approximation of the double-well potential.}
\label{fig:approx_Higgs}
\end{figure}

\begin{figure}[ht]%
\centering
\includegraphics[width=0.85\textwidth]{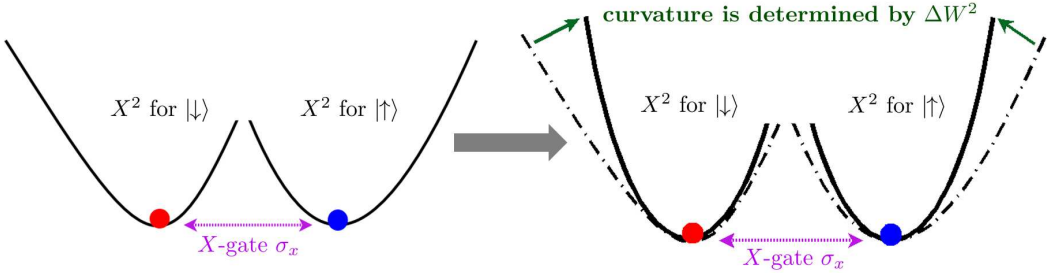}
\caption{Mass-Enhancement in 2-Level-System Approximation.
  In our approximation, the curvature of the wells determines 
  the mass enhancement instead of the radial-direction excitation
  in Fig.\ref{fig:approx_Higgs}.}
\label{fig:mass_enhancement}
\end{figure}

Following the Nambu and Jona-Lasinio's theory \cite{nam61}, and Goldstone's \cite{gol61},
we need the Mexican-hat potential to have a spontaneous symmetry breaking.
Moreover, the Brout-Englert-Higgs mechanism  \cite{eng64, hig64}
requires the Higgs potential, one of the Mexican-hat potentials, 
for the mass generation. 
The interaction of our model does not have the Higgs potential.
It is worthy to emphasize that the Hamiltonian $H$ in Eqs.(\ref{eq:h_XP}) and (\ref{eq:important-limit})
acts on the state space
$\mathbb{C}^{2}\otimes\mathcal{F}_{\mathrm{b}}$. 
Thus, the potential $X^{2}$ in Eq.(\ref{eq:h_XP}) makes a 2-level-system
approximation of a double-well potential (Fig.\ref{fig:approx_Higgs}),
and plays a role as a substitute for the Higgs potential in our story
by employing the $X$-gate $\sigma_{x}$ instead of the
the parity transformation $-X\longleftrightarrow X$.
For the Hamiltonian $H(0, \Omega, 0, 0)$ of the 2-level system coupled to a 1-mode boson, 
the mathematical structure of spontaneous symmetry
breaking for $H(0, \Omega, 0, 0)$
is explained in the last part of 
Section 4 of Ref.\cite{hir15}.
More precisely, the $X$-gate symmetry,
$[ H(0, \Omega, 0, 0), \sigma_{x}]=0$,
makes the global symmetry of our total system,
however, 2-fold degenerate vacuums,
$\mid\downarrow\rangle\otimes{\mid\!\!0}\rangle$ and 
$\mid\uparrow\rangle\otimes\mid{\!\!0}\rangle$,
break $X$-gate invariance,
$\mid\downarrow\rangle\otimes{\mid\!\!0}\rangle\ne
\sigma_{x}\mid\downarrow\rangle\otimes{\mid\!\!0}\rangle
=\mid\uparrow\rangle\otimes{\mid\!\!0}\rangle$,
which usurps the local symmetry
and makes the symmetry breaking on ground state (Fig.\ref{fig:approx_Higgs}).
In the 2-level-system approximation,
the mass enhancement is made by the increment
between the coefficients of $X^{2}$
(Fig.\ref{fig:mass_enhancement}).

The left graph of Fig.\ref{fig:approx_Higgs} shows the schematic image
of the cross section of the Higgs potential with the $XV$-plane.
Here, the variables of the potential are restricted on the real part of $\mathbb{C}$,
and therefore, the Higgs potential of the scalar field, for example,
is given by $V(X)=\mu^{2}X^{2}+\lambda X^{4}+\mu^{4}/(4\lambda)$,
and its value is minimized at $\pm v$ with $v=\sqrt{\frac{-\mu^{2}}{2\lambda}}$,
$\mu^{2}<0$, and $\lambda>0$.
Although the original Higgs potential for the complex scalar field
has the ``global'' $U(1)$-gauge symmetry,
the $U(1)$ transformations are reduced to only the parity transformation,
$-X\longleftrightarrow X$, under the restriction. 
Substituting $X=(H_{\pm}\pm \sqrt{2}v)/\sqrt{2}$ into $V(X)$,
we have $V(X)=\frac{\lambda}{4}H_{\pm}^{4}\pm\sqrt{2}vH_{\pm}^{3}
-\mu^{2}H_{\pm}^{2}$.
Following the Brout-Englert-Higgs mechanism \cite{eng64, hig64},
the mass generation is caused by the excitation in the radial direction
of the Higgs potential in Fig.\ref{fig:approx_Higgs}. 
In particular, the mass generation is determined
by the coefficient of $H_{\pm}^{2}$.
Actually, since the last term should be the mass term
$(m_{H}^{2}/2)H_{\pm}^{2}$, the mass $m_{H}^{\,\,}=\sqrt{-2\mu^{2}}$
in the natural unit is generated \cite{mel17}.  
The right graph of Fig.\ref{fig:approx_Higgs}
is the schematic notion of the 2-level-system
approximation of the double-well potential.
We take the limit, $\lambda\to\infty$, in the potential $V(X)$
keeping $v$ finite, i.e., $-\mu^{2}\sim\lambda$.
Then, we have ${\displaystyle \lim_{\lambda\to\infty}V(0)=\infty}$,
and we reach the broad image of this approximation.
Correspondingly to the coefficient $-\mu^{2}$
of the mass term in $V(X)\sim\frac{\lambda}{4}H_{\pm}^{4}-\mu^{2}H_{\pm}^{2}$
for sufficiently large $\lambda$,
we add the term of $W^{2}$ and increase $C$ in
Eq.(\ref{eq:h_int}) as in Fig.\ref{fig:mass_enhancement} instead.  
In our approximation, therefore,
the mass enhancement is determined by the curvature of the wells
instead of by the radial-direction excitation,
and made by
the increment of $\Delta W^{2}=\left(\frac{2C}{\omega}g(r)^{2}
-\frac{1}{2}\right)W^{2}$ 
in place of the coefficient of $H_{\pm}^{2}$ in Fig.\ref{fig:approx_Higgs}.

We introduce the $1$-mode scalar field $\Phi_{r}$ and its conjugate field $\Pi_{r}$
of the boson getting heavy by
\begin{equation}
\Phi_{r}=\sqrt{\frac{\hbar}{2\omega_{\mathrm{g}}(r)}}\left( B_{r}+B_{r}^{\dagger}\right),\qquad
\Pi_{r}=-i\sqrt{\frac{\hbar\omega_{\mathrm{g}}(r)}{2}}\left( B_{r}-B_{r}^{\dagger}\right),
\label{eq:heavy_r-fields}
\end{equation}
for $0\le r\le 1$.
We denote $\Phi_{1}$ and $\Pi_{1}$ by $\Phi$ and $\Pi$, respectively,
because $B_{1}=B$. 
Then, we have $[\Phi_{r}, \Pi_{r}]=i\hbar$.  
The Lagrangian $L_{r}$ corresponding to $H(r)$ is given by 
$$
L_{r}=\frac{1}{2}\Pi_{r}^{2}-\frac{\omega_{\mathrm{g}}(r)^{2}}{2}\Phi_{r}^{2}
+\frac{\hbar\omega_{\mathrm{a}}(r)}{2}
\left(\mathcal{D}_{-}+\mathcal{D}_{+}\right).
$$
In particular, we have
\begin{align}
L_{1}=\frac{1}{2}\Pi^{2}-\frac{\omega_{\mathrm{g}}^{2}}{2}\Phi^{2} 
\label{eq:L1a}
\end{align}
since $\omega_{\mathrm{a}}(1)=0$.
The Lagrangian $L_{1}$ corresponds to the Hamiltonian $H$
since $B=B_{1}$. 

We introduce a scalar field $\phi$ and
its conjugate field $\pi$ of the light boson by
\begin{align}
\phi=\sqrt{\frac{\hbar}{2\omega}}\left( b+b^{\dagger}\right),\qquad 
\pi=-i\sqrt{\frac{\hbar\omega}{2}}\left( b-b^{\dagger}\right).
\label{eq:light_fields}
\end{align}
We use the fields, $\phi$ and $\pi$, as auxiliary fields for the fields, $\Phi_{r}$ and $\Pi_{r}$. 
Taking the limit $r\to 1$, we have $L_{r}\to L_{1}$.
Thus, using Eqs.(\ref{eq:unitrans1a}), (\ref{eq:heavy_r-fields}), and (\ref{eq:light_fields}),
we can rewrite $L_{r}$ and obtain the limit, 
\begin{align}
  L_{r}
  =&\,\,\frac{1}{2}\pi^{2}-\frac{\omega^{2}}{2}\phi^{2}
  - g(r)\sqrt{2\hbar\omega}\, \sigma_{x}\phi
  -2C\omega g(r)^{2}\phi^{2}
  \nonumber \\
  &-\frac{\hbar g(r)^{2}}{4Cg(r)^{2}+\omega}
  -\,\frac{\hbar\omega_{\mathrm{a}}(r)}{2}\sigma_{z}
  \nonumber \\
\mathop{\longrightarrow}_{r\to 1}&\,\,
L_{1}=\frac{1}{2}\pi^{2}-\frac{\omega_{\mathrm{g}}^{2}}{2}\phi^{2}
  - \mathrm{g}\sqrt{2\hbar\omega}\, \sigma_{x}\phi
-\frac{\hbar \mathrm{g}^{2}}{4C\mathrm{g}^{2}+\omega}.
\label{eq:L1b}
\end{align}
In the Lagrangian $L_{r}$, an extra second-order term, $2C\omega g(r)^{2}\phi^{2}$, appears.
Indeed an effect of $\sigma_{x}$ is invisible in it
since $\sigma_{x}^{2}=1$, but the interaction in the Lagrangian $L_{r}$ is basically
constructed with $\sigma_{x}\phi$ which makes the swap between creation and annihilation
of bosons and the swap between the bosonic and fermionic states.
The increment of the mass enhancement is included in the factor, 
$4C\omega\mathrm{g}^{2}$, in the renormalized frequency $\omega_{\mathrm{g}}$.
Considering the dimension, the mass increment $\Delta m$
is given by $\omega_{\mathrm{g}}=\sqrt{\omega^{2}+(\Delta m)^{2}/\hbar^{2}}$,
that is, $\Delta m=2\sqrt{C\omega}\,\hbar\mathrm{g}$.

We here summarize the above results.
1) The 2-level-system approximation works instead of the Higgs potential,
and then, the transition from $\mathcal{N}=2$ SUSY to its spontaneous breaking
takes place. 
2) The transition changes the free field $\phi$ of the light boson
to the free field $\Phi$ of heavy boson.
The heavy boson acquires a part of its mass from the excitation of
the light boson then, caused by the $A^{2}$-term.
This makes the mass enhancement for the fermionic states
as well as for the bosonic ones. 
3) The Lagrangian $L_{1}$ has the spin-chiral symmetry,
$[\sigma_{x}, L_{1}]=0$, 
though the Lagrangian $L_{r}$ does not have it,
$[\sigma_{x}, L_{r}]\ne 0$, for $0\le r<1$
because of the existence of the spin term,
$-\hbar\omega_{\mathrm{a}}(r)\sigma_{z}/2$.

We can restate the results in terms of Hamiltonian.
The transition from the Hamiltonian $H(0)=H_{\mbox{\tiny SS}}$ of the light boson 
to the Hamiltonian $H(1)=H$ of the heavy boson is obtained:
\begin{align}
H(0)&=H_{\mbox{\tiny SS}}=\hbar\omega\left( b^{\dagger}b+\frac{1}{2}\right)
+\frac{\hbar\omega}{2}\sigma_{z}
\nonumber \\ 
&\Longrightarrow
H(1)=H=\hbar\omega_{\mathrm{g}}\left( B^{\dagger}B+\frac{1}{2}\right),
\label{eq:transition}
\end{align}
where we omit the 2-by-2 identity matrix $1$. 
According to the facts in Section \ref{sec2},
Eq.(\ref{eq:transition}) says that 
the transition brings the $\mathcal{N}=2$\, SUSY Hamiltonian $H(0)$ to
its spontaneous-breaking Hamiltonian $H(1)$, 
and the transition yields the mass enhancement with the increment
$\Delta m=2\sqrt{C\omega}\,\hbar\mathrm{g}$,
determined by $\omega_{\mathrm{g}}^{2}=\omega^{2}+(\Delta m)^{2}/\hbar^{2}$
coming from the increment of the mass term, $-\left((\Delta m)^{2}/(2\hbar^{2})\right)\phi^{2}$. 
It is worthy to note again that Cai \textit{et al.} report
the observation of the transition, Eq.(\ref{eq:transition}),
in the case $C=0$ \cite{cai22}.

\begin{figure}[ht]%
\centering
\includegraphics[width=0.45\textwidth]{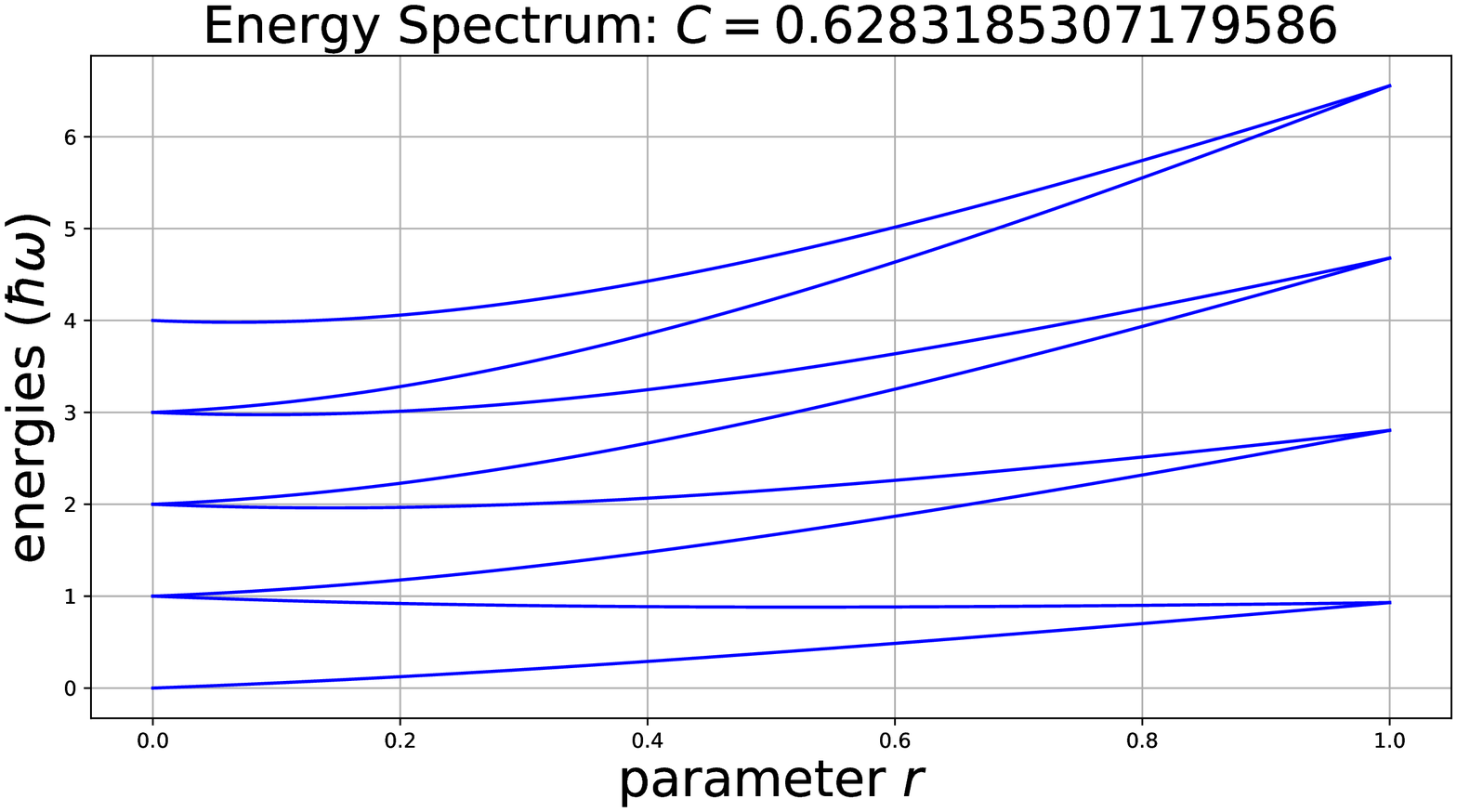}
\quad
\includegraphics[width=0.45\textwidth]{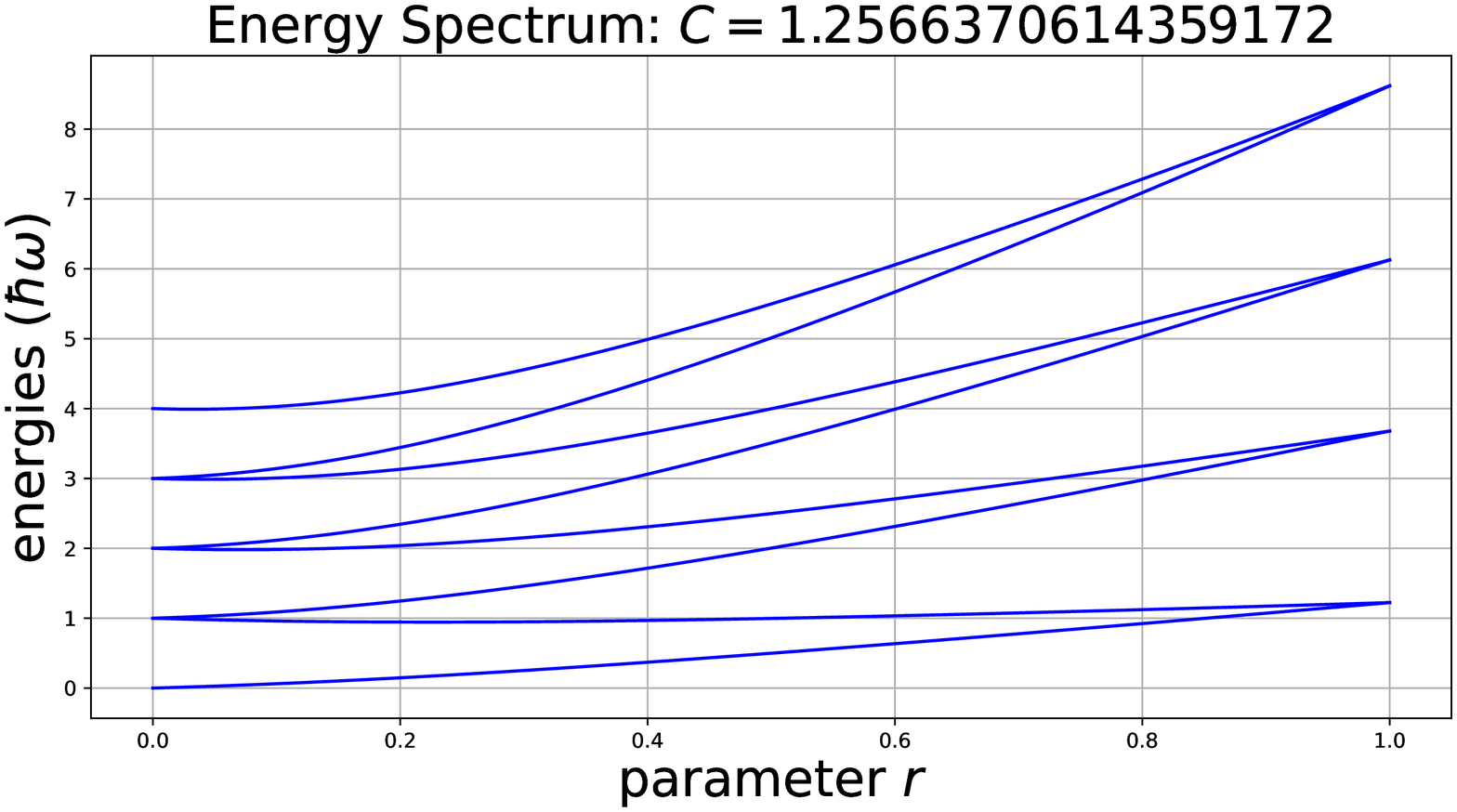}
\vspace*{17mm}
\caption{Mass Enhancement in Energy Spectrum of
  $H(r)$ with $\omega=6.2832$ and $\mathrm{g}=6.2832$. 
  A ground state energy and six excited state energies
  from the bottom are shown in each graph.
  The graphs respectively show the energy spectrum for 
  $C=0.628$ and $C=1.257$ from the left.
    In these numerical calculations, $\omega_{\mathrm{a}}(r)=(1-r)\omega$
    and $g(r)=r\mathrm{g}$ are employed.}\label{fig:Fig_1}
\end{figure}

Since each energy level of $H(r)$ is guaranteed for its convergence
as $r\to 1$ by the limit in the norm resolvent sense (see Theorem VIII.24 of \cite{rs1}),
we are interested in the energy spectrum of $H(r)$ for every $r$ with $0\le r\le 1$.
Fig.\ref{fig:Fig_1} shows its two examples by numerical calculations with QuTiP \cite{nori1, nori2}.

\section{Conclusion and discussion}\label{sec4}

We have proposed a mathematical model, though very simple,
for quantum simulation of a mass enhancement in the SUSY breaking.
This model is based on the quantum Rabi model with the $A^{2}$-term,
and reveals a transition from the $\mathcal{N}=2$\, SUSY to its spontaneous breaking.
We have proved that the $A^{2}$-term works
for the mass enhancement in the fermionic states as well as in the bosonic ones.
We have shown that, in the process of the transition, 
the quasi-particle of the light bosons eats the effect of the $X$-gate
and becomes the heavy boson.

We have explained that the qubit system (i.e., the 2-level system)
coupled with boson is good at simulating the so-called double-well potential
such as the Higgs potential.
For another example, we know that the quantum Rabi model has some properties
similar to the instanton \cite{col77, CC77}
as well as the spin-boson model (see \cite[Theorem 1.5]{hir99} and
\cite[Appendix B]{hir11}).
It is known that the instanton gives a non-perturbative effect in $\mathcal{N}=2$
SUSY gauge field theory \cite{sei88}. 
A quantum simulation of Weinberg-Salam theory might be simulated 
using a qubit system coupled with boson such as the microwave photon
\cite{sav14, sav22}. 

In the case without the $A^{2}$-term, it is reported that 
the transition is experimentally observed in a trapped ion quantum simulator by Cai \textit{et al.} \cite{cai22}.
Thus, a future experimental problem would be whether $A^{2}$-term can be added to
their experimental set-ups in a quantum simulator,
and an experimental observation of the energy spectrum can be performed.

\begin{figure}[ht]%
\centering
\includegraphics[width=0.3\textwidth]{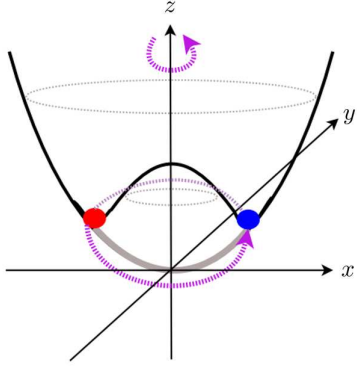}
\caption{Global $U(1)$ Symmetry of Higgs Potential.
  The graph shows the schematic image
  of the Higgs potential among the Mexican-hat potentials.
  The spontaneous breaking of the symmetry for the rotation around
  the $z$-axis is continuous, and produces the Nambu-Goldstone bosons
  following Nambu and Jona-Lasinio's theory \cite{nam61},
  and Goldstone's \cite{gol61}.} 
\label{fig:approx_Higgs2}
\end{figure}

The results in this paper raise the following issues: 
Can we see a fingerprint of the mode of the so-called Goldstino
(i.e., Nambu-Goldstone fermion) \cite{sal74, wit82, bin06, bau15a, bau15b, san16, san17, bla17, ma21, taj21}
in the SUSY breaking for our quantum-mechanical model?
The Higgs potential makes the continuous symmetry with respect to
the rotation around the $z$-axis in Fig.\ref{fig:approx_Higgs2},
and then, the Nambu-Goldstone bosons appear
for the degree of freedom for $\arg(x+iy)$ in Fig.\ref{fig:approx_Higgs2}.
On the other hand, SUSY is discontinuous, and ours is $X$-gate symmetry.
Can a fingerprint of Goldstino be detected even in such discrete symmetry? 
If we can grasp the Goldstino's influence,
what is the mathematical characterization 
between the Goldstino and the supercharges in our model?
From this point of view, it is worthy to note that Cai \textit{et al.} 
have been developing the technology to observe the supercharges \cite{cai22}. 
In our SUSY breaking, the oscillation between the bosonic and fermionic states
is that between qubits, the down-spin and the up-spin states,
with the same boson number $n$.
Is there any relation between the Goldstino and the Rabi oscillation?

\begin{acknowledgments}
  The author acknowledges the support from
  JSPS Grant-in-Aid for Scientific Researchers (C) 20K03768.
  He wishes to thank Masahiko Ichimura for his comment.
  He would like to dedicate this study to Hiroshi Ezawa and 
  Elliott H. Lieb on the occasions of their 90th birthdays. 
\end{acknowledgments}

\bibliography{hirokawa_rev3}

\end{document}